\documentclass{aa}
\usepackage[varg]{txfonts}
\usepackage{graphicx}  
\usepackage{amsmath}   
\usepackage{subfigure}
\usepackage{epstopdf}
\usepackage[colorlinks, linkcolor=blue, urlcolor=blue, citecolor=blue]{hyperref}
\usepackage{booktabs}
\usepackage{ulem}
\usepackage{threeparttable}
\usepackage{enumitem}
\usepackage{longtable}
\def\be{\begin{equation}}
\def\ee{\end{equation}}
\usepackage{CJK}
\usepackage{upgreek}

\begin{document}

\title{FRB 20200428 and potentially associated hard X-ray bursts: Maser emission and synchrotron radiation of electrons in a weakly magnetized plasma?}

\begin{CJK*}{UTF8}{gbsn}
\author{Xiao Li \inst{\ref{inst1}}
\and En-Wei Liang\inst{\ref{inst1}}
}

\institute{$^1$Guangxi Key Laboratory for Relativistic Astrophysics, School of Physical Science and Technology, Guangxi University, Nanning 530004, China\label{inst1}
\\
\email{lew@gxu.edu.cn}
}

\date{Received XXX / Accepted XXX}

\abstract{
The temporal and spatial coincidence between FRB 20200428 and hard peaks of the X-ray burst from SGR 1935+2154 suggests their potential association. 
We attributed them to the plasma synchrotron maser emission and synchrotron radiation of electrons in weakly magnetized, relativistically moving plasma blobs, and Monte Carlo simulation analysis shows that our model can predict observable fast radio burst outbursts and associated hard X-ray bursts with current telescopes. We constrained the properties of the blobs, including the Lorentz factor $\Gamma=5-30$, the magnetization factor $\sigma=6\times10^{-5}\sim 2\times 10^{-4}$, the electron Lorentz factor $\gamma_{\rm e,s}=(1.8-3.3)\times10^4$, and the plasma frequency $\nu_P=2.48 -42.61$ MHz. The inferred size of the blobs is $\sim 10^{9-10}$ cm, and it is located $\sim 10^{12-14}$ cm from the central engine. By adopting fine-tuned parameter sets, the observed spectra of both the FRB 20200428 outbursts and X-ray bursts can be well represented. The peak flux density ($F_{\rm\nu_{ pk}}$) of plasma maser emission is sensitive to $\sigma$ and $\nu_P$. Variation in $F_{\rm \nu_{pk}}$ can be more than 10 orders of magnitude, while the flux density of the synchrotron emission only varies by $1-2$ orders of magnitude. This can account for the observed sub-energetic radio bursts or giant radio pulses from SGR 1935+2154.}


\titlerunning{Maser and synchrotron radiations as mechanisms for FRB 20200428 and associated X-ray bursts}
\authorrunning{Xiao Li, et al.}

\maketitle

\section{Introduction}\label{sec:intro}
\end{CJK*}
\noindent Fast radio bursts (FRBs) are bright, short-duration radio transients \citep{2007Sci...318..777L,2019ARA&A..57..417C,2022A&ARv..30....2P}. Most FRBs are extragalactic events with a typical dispersion measurement of $\sim 88- 3038\, {\rm {pc}}\,{\rm {cm^{-3}}}$ \citep{2021ApJ...910L..18B,2021ApJS..257...59C}. This has been confirmed through the identification of the host galaxies of some FRBs \citep{2020ApJ...903..152H,2022AJ....163...69B}. To date, over 800 FRBs have been detected \citep{2016PASA...33...45P,2021ApJS..257...59C}, more than 60  of which exhibit repetitive behaviors 
\citep{2020ApJ...891L...6F,2023ApJ...947...83C}\footnote{\url{https://blinkverse.alkaidos.cn}}. The high brightness temperature ($\rm {T_B} \geq {10^{35}} K$) of FRBs suggests that their radiation mechanism must be coherent \citep{2020Natur.587...45Z,2021Univ....7...56L,2021SCPMA..6449501X}. Various coherent mechanisms have been proposed to interpret the emissions of FRBs, such as synchrotron maser radiation in relativistic shocks under strong magnetization conditions \citep{2014MNRAS.442L...9L,2017ApJ...843L..26B,2020ApJ...896..142B,2019MNRAS.485.4091M}, plasma synchrotron maser radiation under weak magnetization conditions \citep{2017ApJ...842...34W,2025A&A...695A.100L}, and coherent curvature radiation by nearby bunches in the magnetosphere \citep{2014PhRvD..89j3009K,2017MNRAS.468.2726K,2018ApJ...868...31Y}. The origin of FRBs remains enigmatic and is a subject of hot debate (see \citealt{2019PhR...821....1P,2023RvMP...95c5005Z} for reviews); most hypotheses involve compact objects that have been modeled according to the energetics and temporal variability of FRBs. Magnetars, highly magnetized young neutron stars, have been widely discussed as a potential progenitor of FRBs in the literature \citep{2014MNRAS.442L...9L,2018ApJ...868...31Y,2018ApJ...868L...4M,2019MNRAS.485.4091M,2017ApJ...843L..26B,2020ApJ...896..142B}.
The detection of magnetar SGR 1935+2154 in the Milky Way associated with FRB 20200428 in time and space hints that at least some FRBs originate from magnetars \citep{2020Natur.587...59B,2020Natur.587...54C}.\par

FRB 20200428 is the only FRB source that has been detected in the Milky Way to date; it has a dispersion measurement of $332.7\; \rm{pc\,cm^{-3}} $ and a rotation measure of $\sim 116 \;\rm {rad\; m^{-2}}$ \citep{2020Natur.582..351C,2020Natur.587...59B}. Three bursts from FRB 20200428 have been detected. The first pulse was detected only by Canadian Hydrogen Intensity Mapping Experiment (CHIME; $400-800$ MHz), and the second was detected simultaneously by CHIME and the Survey for Transient Astronomical Radio Emission 2 (STARE2; $1.28-1.47$ GHz). 
Excitingly,
a hard X-ray burst was detected from SGR 1935+2154 by several instruments, including Insight-Hard X-ray Modulation Telescope (Insight-HXMT), INTErnational Gamma-Ray Astrophysics Laboratory (INTEGRAL), Astro rivelatore Gamma a Immagini Leggero (AGILE), and Konus-Wind (KW) during the FRB burst \citep{2021NatAs...5..378L,2020ApJ...898L..29M,2021NatAs...5..401T,2021NatAs...5..372R}. 
The burst showed two narrow peaks with a separation of $\sim$ 30 ms, consistent with the separation between the two bursts in FRB 20200428 detected by CHIME \citep{2020Natur.582..351C,2021NatAs...5..378L,2021NatAs...5..372R}. 
The dispersion delay between the double radio and narrow X-ray peaks is fully consistent with that of FRB 20200428, which indicates that the two peaks in X-ray and radio emission most likely have a common origin \citep{2021NatAs...5..378L,2021NatAs...5..372R}.
\par

The association between FRB and X-ray may provide insight into their origin and mechanism. Many theoretical models have been proposed to explain this unique FRB--X-ray event. Some are based on the idea that FRBs are produced by coherent curvature radiation (e.g., \citealt{2020ApJ...897L..40D,2020MNRAS.498.1397L,2020ApJ...901L..13Y,2020ApJ...899..109W,2020ApJ...898L..55G}), while others are based on synchrotron maser radiation (e.g., \citealt{2020ApJ...904L...5X,2020ApJ...899L..27M,2020ApJ...900L..26W,2021MNRAS.500.2704Y}). Motivated by the simultaneous observations of both FRB 20200428 and two narrow X-ray peak bursts from Galactic magnetar SGR 1935+2154, we investigated whether they arise from plasma synchrotron maser emission and synchrotron radiation of electrons in weakly magnetized relativistic plasma blobs. The paper is organized as follows. Our model is outlined in Sect. \ref{sec:Mod}, and the numerical calculation results are presented in Sect. \ref{sec: num cal}. The discussion and conclusions are given in Sects. \ref{DIS} and \ref{CON}, respectively. Throughout, we adopt a flat $\Lambda$ cold dark matter universe with cosmological parameters $H_{0}$=67.7$\mathrm{~km}$ $\mathrm{~s}^{-1}$ $\mathrm{Mpc}^{-1}$ and $\Omega_{m}=0.31$ \citep{2016A&A...594A..13P}. 

\section{Model}\label{sec:Mod}
We propose that FRB 20200428 and its associated X-ray peak bursts are generated by plasma synchrotron maser emission and the synchrotron radiation of electrons in weakly magnetized relativistic plasma blobs, which are induced by plasma instabilities triggered by the injected ejecta from the central engine (e.g., \citealt{2025A&A...695A.100L}). The emitting plasma blobs move toward observers with a bulk Lorentz factor of $\Gamma$, and the electrons are randomly accelerated. The plasma blobs have a magnetization factor of $\sigma \ll 1$, which is defined as $\sigma={{ {B^2}}}/{{4\pi {m_{\rm e}}{c^2}\gamma_{\rm e} n_{\rm e}}} = {\left( {{{{\nu _B}}}/{{{\nu_{ P}}}}} \right)^2}$, where ${\nu_{ P}} = {\left( {{{n_{\rm e}{{\rm e}^2}}}/{{\pi \gamma_{\rm e} {m_{\rm e}}}}} \right)^{1/2}}$ is plasma frequency and ${\nu _B} = {\rm{e}}B/2\pi {\gamma _{\rm{e}}}{m_{\rm{e}}}c$ is the cyclotron frequency of the relativistic electron, $n_{\rm e}$ is the relativistic electron number density, $\rm e$ is the electron charge, $m_{\rm e}$ is the electron rest mass, $B$ is the magnetic field, and $\gamma_{\rm e}$ is the electron Lorentz factor \citep{2021Univ....7...56L}. It has been shown that plasma synchrotron maser emission can be generated in a weakly magnetized relativistic plasma created by collisionless shocks (e.g., \citealt{2002ApJ...574..861S,2017ApJ...842...34W,2019ApJ...875..126G,2025A&A...695A.100L}). 
The FRB burst is attributed to the plasma synchrotron maser emission, while the synchrotron emission of the electrons accounts for the X-ray burst.
\par

It is worth noting that the power per unit frequency emitted by a single relativistic electron with Lorentz factor $\gamma_{\rm{e}}$ in a polarization mode ${\left[ \perp, \| \right]}$ within such a plasma blob is modified to 
\citep{1989aetp.book.....G,2002ApJ...574..861S}
\begin{equation}\label{sec2 equ1}
\begin{aligned}
P_\nu ^{\left[ \perp, \| \right]}\left( {{\gamma _{\text{e}}}} \right)=&\frac{\sqrt{3} {\rm {e^3}} { B}\sin \chi}{2 m_{\rm e} c^{2}}\left[{1 + \gamma _{\rm{e}}^2\left( {1 - {{\rm n}^2}} \right)}\right]^{-1 / 2} \frac{\nu}{\tilde{\nu}_{\rm c}} \times\\ &\left[\int_{\nu / \tilde{\nu}_{\rm c}}^{\infty} K_{5 / 3}(z) d z \pm K_{2 / 3}\left(\frac{\nu}{\tilde{\nu}_{\rm c}}\right)\right] \;,
\end{aligned}
\end{equation} where $\bot$ and $\parallel$ denote the linear polarization perpendicular and parallel to the projection of the magnetic field on the plane of observation, respectively. $\chi$ is the pitch angle, $\rm n$ is the refractive index, $K_{5/3}$ and $K_{2/3}$ are the modified Bessel functions, and \begin{equation}\label{sec2 equ2}{\tilde \nu }_{\rm{c}} = \frac{{3{\rm{e}}B\sin \chi }}{{4\pi {m_{\rm{e}}}c}}\gamma _{\rm{e}}^2{\left[ {1 + \gamma _{\rm{e}}^2\left( {1 - {{\rm n}^2}} \right)} \right]^{ - 3/2}} \;.
\end{equation}
The refractive index in a relativistic plasma depends on the energy and angular distribution of particles \citep{1984oep.....9.2444A}. For a monoenergetic distribution of electrons in a weakly magnetized relativistic plasma, we have ${\rm {n^2}} = 1 - (\frac{\nu_P}{\nu})^2$ \citep{2002ApJ...574..861S}.
\par

We assumed that the electrons in each plasma blob follow an isotropic monoenergetic distribution. Therefore, the radiation intensity for an individual blob in 
polarization mode $\bot$ is given by (e.g., \citealp{2002ApJ...574..861S}) 
\begin{equation}\label{sec2 equ3}
 I_\nu = j_\nu \Delta \frac{{1 - {e^{ - \tau _\nu }}}}{{\tau _\nu }}\;,
\end{equation}
where \begin{equation}\label{sec2 equ4}
j_\nu = \int {\frac{{P_\nu \left( {{\gamma _{\rm{e}}}} \right)}}{{4\pi }}\frac{{d{n_{\rm{e}}}}}{{d{\gamma _{\rm{e}}}}}} d{\gamma _{\rm{e}}}
\end{equation} 
is the specific emissivity, ${\tau _\nu } = \alpha _\nu \Delta $ is the optical depth, $\Delta$ is the width of the radiating region along the line of sight, and \begin{equation}\label{sec2 equ5}
 \alpha _\nu = - \frac{1}{{4\pi {m_{\rm{e}}}{\nu ^2}}}\int {\gamma _{\rm{e}}^2} P_\nu \left( {{\gamma _{\rm{e}}}} \right)\frac{d}{{d{\gamma _{\rm{e}}}}}\left( {\gamma _{\rm{e}}^{ - 2}\frac{{d{n_{\rm{e}}}}}{{d{\gamma _{\rm{e}}}}}} \right)d{\gamma _{\rm{e}}}\;
\end{equation} is the synchrotron self-absorption coefficient obtained via the Einstein coefficient method
\citep{1989aetp.book.....G,2002ApJ...574..861S}. 
In the case of the electron distribution taken as a delta function, $\frac{{d{n_{\rm{e}}}}}{{d{\gamma _{\rm{e}}}}} = \delta \left( {\gamma _{\rm e} - {\gamma _{\rm e,s}}} \right)$, the synchrotron self-absorption coefficient can be estimated as \citep{2025A&A...695A.100L}
\begin{equation}\label{sec2 equ6}
 \alpha _\nu (g,y) = 2{\alpha _0}{y^{ - 3}}\left[ {{f}(x) + \left( {\frac{1}{2} - \frac{{{y^2}}}{g}} \right)x{f^{\prime }}(x)} \right]
\end{equation}
with \begin{equation}\label{sec2 equ7}
 {\alpha _0} = \frac{{\pi {\nu _P}}}{{2\sqrt 3 c}}{\sigma^{3/4}} \sin \chi \;,
\end{equation}
and the specific emissivity can be estimated as
\begin{equation}\label{sec2 equ8}
j_\nu (g,y) = {j_0}{(1 + \frac{g}{{{y^2}}})^{ - 1/2}}g\left[ {x{f}\left( x \right)} \right]
\end{equation} 
with \begin{equation}\label{sec2 equ9}
 {j_0} = \frac{{\pi {m_e}}}{{2\sqrt 3 c}}\nu _P^3\sin \chi \;,
\end{equation}
where $g =\gamma _{{\text{e,s}} }^2\sigma^{1/2} $, $y = \frac{\nu }{{\nu _{\rm R}^*}}$, $\nu _{\rm R}^* = \sigma^{-1/4} {\nu_{ P}}$, $f(x) = \int_x^\infty {{K_{5/3}}} (z)dz$, and ${x = y{{\left( {{g^{ - 1}} + {y^{ - 2}}} \right)}^{3/2}}/\sin \chi }$. Therefore, the radiation flux density of the individual blob in the observer's frame can be estimated as \begin{equation}\label{sec2 equ10}
{F_\nu }\left( {{\nu _{{\rm{obs}}}}} \right) = \frac{{(1 + z){\Gamma ^3}j{{_{{\nu ^\prime }}^\prime }}\left( {{\nu ^\prime }} \right)\frac{{1 - {e^{ - \tau {{_{{\nu ^\prime }}^\prime }}\left( {{\nu ^\prime }} \right)}}}}{{\tau {{_{{\nu ^\prime }}^\prime }}\left( {{\nu ^\prime }} \right)}}{V^\prime }}}{{D_{\rm{L}}^2}}\;,
\end{equation} 
where the prime means that the corresponding quantities are measured in the comoving frame, $z$ is the redshift, and $D_{\rm L}$ is the luminosity distance. The observed peak frequency can be estimated as \citep{2025A&A...695A.100L}
\begin{equation}\label{sec2 equ11}
 {\nu _{\rm{pk}}} = 0.70{\rm Hz}~(1 + z)^{-1}\Gamma {\sigma ^{ - 1/4}}{\nu _{{P}}} = 0.70{\rm{GHz}}~{(1 + z)^{ - 1}}{\Gamma _2}\sigma _{ - 4}^{ - 1/4}{\nu _{{{P}},{\rm{6}}}},
\end{equation} where the notation $Q_n=Q/10^{n}$ is adopted in cgs units. The frequency range of plasma synchrotron maser emission and the peak frequency depend on $\gamma _{{\rm{e,s}} }^2{\sigma ^{1/2}}$. As shown in \cite{2025A&A...695A.100L}, if 
$\gamma _{{\rm{e,s}} }^2{\sigma ^{1/2}}>50$, the plasma synchrotron maser emission regime in the observer's frame is confined to the frequency range $0.4\Gamma\nu _{\rm R^*} < \nu < \Gamma\nu _{\rm R^*}$.

\section{Numerical calculation results}\label{sec: num cal}

\begin{figure}
\includegraphics[width=1\columnwidth]{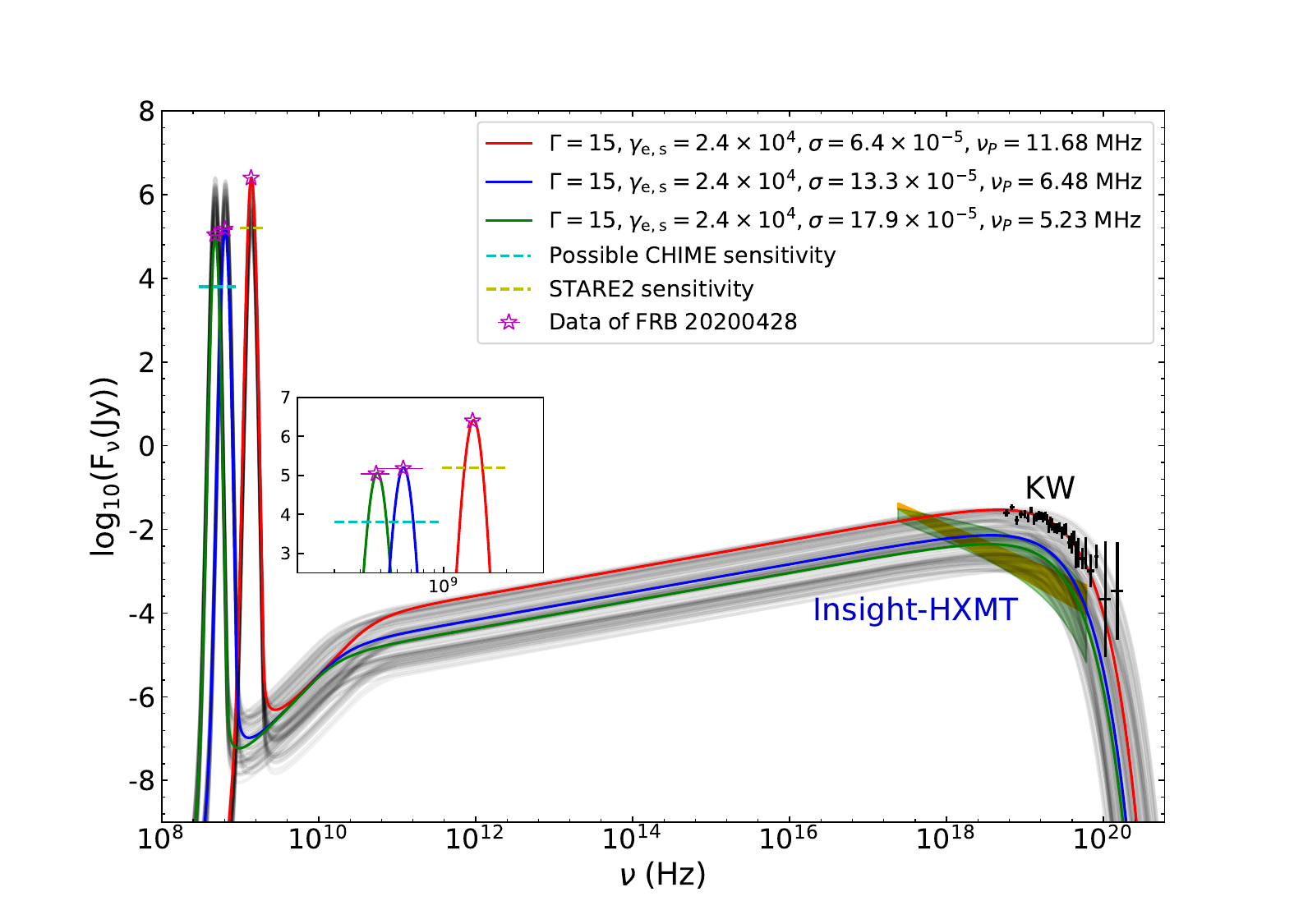}
 \centering
\caption{Observed peak frequencies and corresponding peak flux densities of FRB 20200428 (stars) and the spectra of potentially associated X-ray bursts from SGR 1935+2154 observed with Insight-HXMT (green and orange bow ties) and KW (black data points) together with our model analysis results. The gray lines represent the model-predicted spectra of both FRB and associated X-ray bursts derived from 100 randomly selected model parameter sets based on our Monte Carlo simulation analysis and assuming $\chi=\frac{\pi }{4}$. The red, blue, and green lines represent the model fit to the observed spectra of three outbursts of FRB 20200428 together with potentially associated X-ray bursts from SGR 1935+2154.}\label{MyFig1}
\end{figure}

The CHIME telescope detected two bursts from FRB 20200428. Their temporal separation is $\sim$ 30 ms. The first burst was detected in the frequency range $400-550$ MHz, and its peak flux density is 110 kJy. The second burst was detected in the frequency range $500-800$ MHz with a flux density of 150 kJy \citep{2020Natur.582..351C}. The STARE2 telescope detected a burst in the frequency range $1290-1468$ MHz almost simultaneously to the second burst observed with CHIME. In our analysis, we considered the two detections to be independent bursts\footnote{
One may have doubts as to whether the bursts detected by the two telescopes are the same event. We note that the spectrum of the second burst of FRB 20200428 detected by CHIME is narrowly banded and can be fitted with a Gaussian function \citep{2020Natur.582..351C}.
However, the spectrum observed by STARE2 cuts off at $\sim 1.29$ GHz (see Fig.1 in \citealt{2020Natur.587...59B}). Therefore, the spectrum observed by CHIME should not be an extension of the burst detected by STARE2 since
STARE2 covers a frequency range of $1.28-1.47$ GHz but CHIME covers $0.4-0.8$ GHz. In addition, the burst detected by CHIME lasts 0.335 ms, whereas the burst detected by STARE2 has a temporal width of 0.61 ms \citep{2020Natur.582..351C,2020Natur.587...59B}. Thus, the bursts detected by the two telescopes are likely two separate events from two blobs.}.
The peak flux density of the pulse detected by STARE2 was approximately $2.45$ MJy \citep{2020Natur.587...59B}. The spectrum of the X-ray bursts from SGR 1935+2154 that may be associated with FRB
20200428 was derived jointly using Insight-HXMT high-energy, medium-energy, and low-energy data ($1-250$ keV; \citealt{2021NatAs...5..378L}). The hard X-ray bursts detected by KW were in the $20-500$ keV range \citep{2021NatAs...5..372R}, as shown in Fig. \ref{MyFig1}.

\begin{figure}
\includegraphics[width=0.48\linewidth]{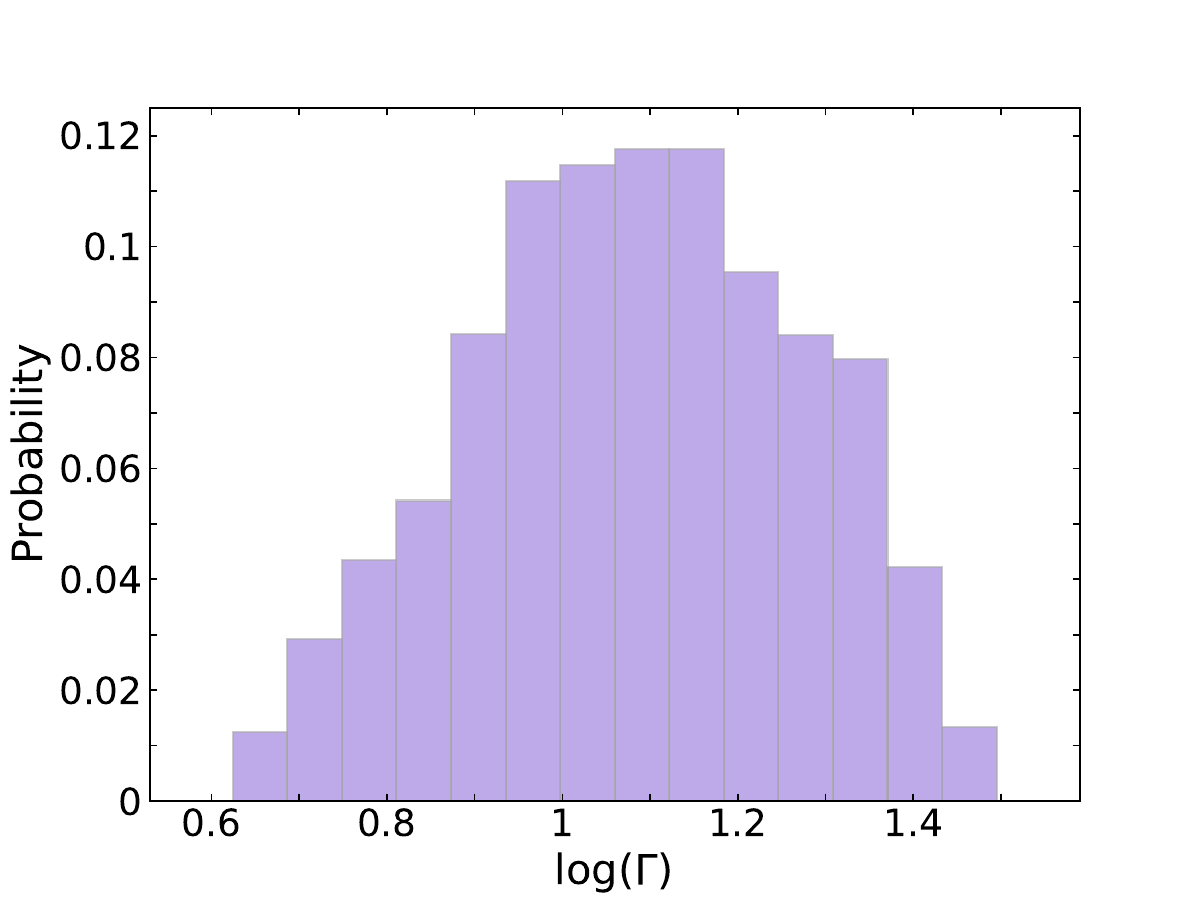}
\includegraphics[width=0.48\linewidth]{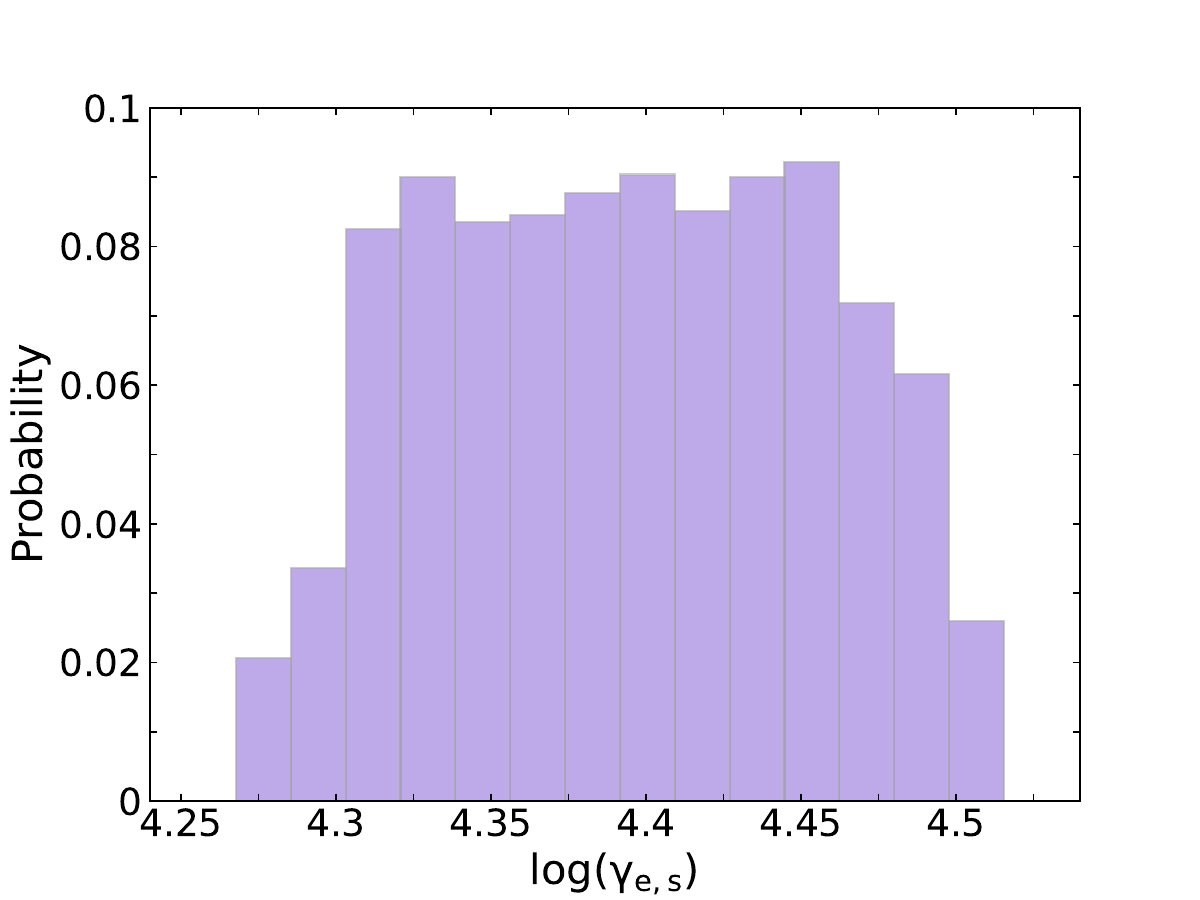}
\includegraphics[width=0.48\linewidth]{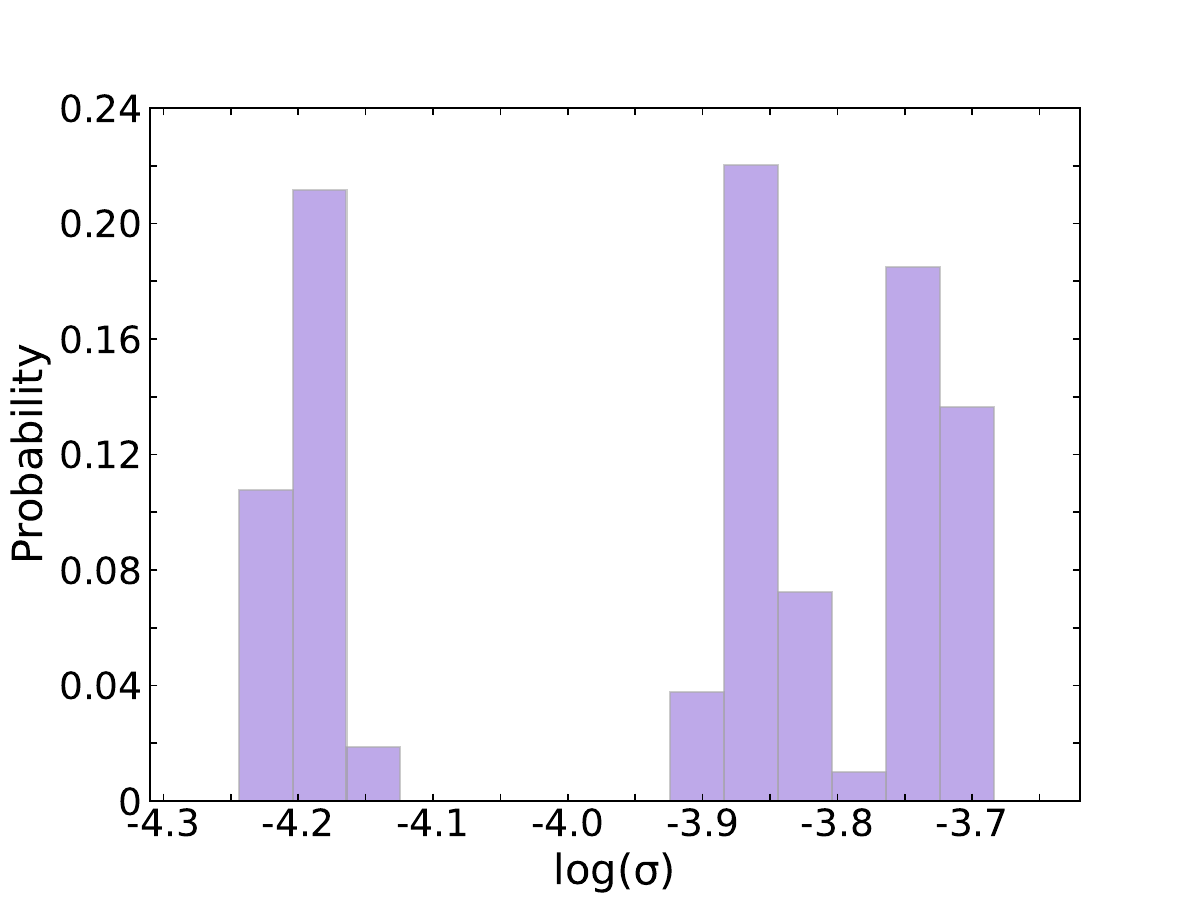}
\includegraphics[width=0.48\linewidth]{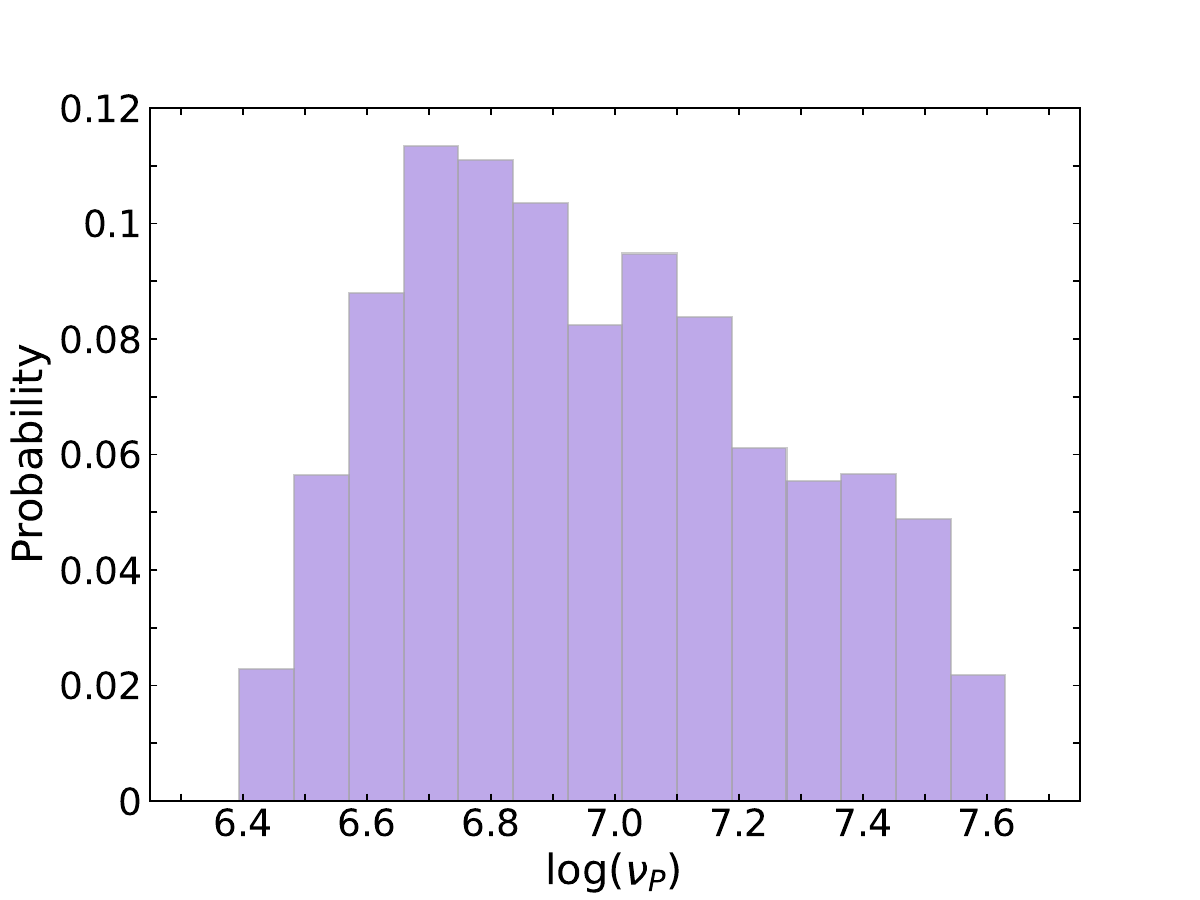}
\centering
\caption{Probability 
histograms of $\Gamma$, $\sigma$, $\gamma_{\rm e,s}$, and $ \nu_P$ derived from our simulation analysis.}\label{MyFig2}
\end{figure}

We calculated the plasma synchrotron maser and synchrotron emission of the electrons in an individual plasma blob. Through a Monte Carlo simulation analysis, we sampled model parameter sets of $\{{{\Gamma,\gamma _{{\rm{e,s}} }},\sigma ,{\nu_{ P}} }\}$, which can produce observable FRB outbursts with CHIME and STARE2, and their associated X-ray bursts from SGR 1935 + 2154 with Insight-HXMT
and KW. Following \cite{2025A&A...695A.100L}, the size of the individual blob in the comoving frame was estimated as $\Delta=\Gamma c \delta t=3\times 10^{7}\Gamma$ cm, assuming $\delta t=1$ ms, where $c$ is the light speed. The luminosity distance to SGR 1935+2154 is estimated to be in the range $6.6-12.5$ kpc \citep{2018ApJ...852...54K,2020ApJ...905...99Z,2020ApJ...898L...5Z}. We set a fiducial value of 10 kpc ($z\sim0$).
The procedure of our Monte Carlo simulation is described below.
\begin{enumerate}
\item We fixed the $\nu _{{\rm{pk}}}$ value at 0.475, 0.64, or 1.37 GHz, as observed by CHIME and STARE2, and generated a set of $\{\Gamma, \gamma_{\rm {e,s}}, \sigma\}$ values assuming that they are uniformly distributed in the ranges $\Gamma \in [1,200]$, $\gamma_{\rm {e,s}}\in[10^3,{10^5}]$, 
and $\sigma\in [\gamma _{{\rm{e,s}} }^{ - 4} ,1)$. The distribution range of $\sigma$ was derived from the weak magnetization condition ($\gamma _{\rm {e,s}}^2 > {\sigma^{-1/2}} > 1$). 
\item We calculated the plasma synchrotron maser emission of the electrons in the blob. We first calculated the $\nu_{ P}$ value with Eq. \eqref{sec2 equ11} for the case of $\gamma _{{\rm{e,s}} }^2{\sigma ^{1/2}}>50$ \citep{2025A&A...695A.100L}, and then calculated the simulated peak flux density ($F^{\rm sim}_{\rm \nu_{pk}}$) at $\nu_{{\rm{pk}}}$ with Eq. \eqref{sec2 equ10} for the parameter set of $\{{{\Gamma,\gamma _{{\rm{e,s}} }},\sigma ,{\nu_{ P}} }\}$. We checked whether the $F^{\rm sim}_{\rm \nu_{pk}}$ value is in the range $0.11 < F_{\rm \nu}^{\rm sim} < 2.5$ MJy, as observed with CHIME and STARE2. If it was, we moved on the next step. Otherwise, we went back to Step 1. 
\item We calculated the peak frequency of the synchrotron emission of the electrons in an attempt to explain the associated X-ray bursts with 
$\nu=0.45\Gamma^2\gamma _{\rm{e,s}}^3 \nu_B$ 
and calculated corresponding flux density ($F_{\nu}$) using Eq. \eqref{sec2 equ10}. 
We checked whether $\nu\in[2, 8] \times10^{18}\; {\rm GHz}$ and $F_{ \nu}\in [10^{-3}, 10^{-1.5}]\;{\rm Jy,}$ as observed with Insight-HXMT and KW. If they were, the parameter set was selected as one possible model parameter set. Otherwise, we discarded this parameter set and repeated the above steps. 
\end{enumerate} 
We repeated this procedure to generate a sample of $5000$ sets of model parameters.\ We show their probability histograms in Fig. \ref{MyFig2}. The probability distribution of $\Gamma$ is found to typically range from 5 to 30 with a median of 13. The probability distribution of $\gamma_{\rm e,s}$ is uniformly distributed in the range $1.8\times10^4-3.3\times10^4$. The distribution of $\sigma$ has three separated peaks, at $\rm log(\sigma)=\left\{-4.21,-3.95,-3.88\right\}$, which correspond to the three observed FRB outbursts with different $\nu_{\rm pk}$ values. 
The distribution of $\nu_P$ ranges from $\log \nu_{P}=6.40$ ($\nu_{P}=2.48 \,{\rm MHz}$) to $\log \nu_{ P}=7.6$ ($\nu_{ P}=42.61\, {\rm MHz}$).

We randomly selected 100 sets of model parameters from the 5000 sets and illustrate the spectra derived from our model for these 100 in Fig. \ref{MyFig1}. The predicted FRB outbursts and their associated X-ray bursts are detectable. By further fine-tuning the parameters as $\{{{\Gamma,\gamma _{{\rm{e,s}} }},\sigma ,{\nu_{ P}/{\rm MHz}} }\}=\{15,2.4\times10^4,6.4\times10^{-5},11.68\}, \{15,2.4\times10^4,13.3\times10^{-5},6.48 \}$, and $\{15,2.4\times10^4,17.9\times10^{-5},5.23\}$, we obtain ${\nu} _{\rm pk}=1.37,0.63,0.47\;{\rm {GHz}}$ and ${F_{{\nu _{\rm pk}}}} = 2.5,0.15,0.11$ MJy, as shown by the solid red, blue, and green lines in Fig. \ref{MyFig1}. The predicted spectra closely fit the three observed FRB 20200428 bursts and potentially associated X-ray bursts.
The synchrotron emission of the electrons is in the range$10^{10}\sim 10^{20}$ Hz. It peaks around the hard X-ray band, with a flux density of $10^{-5}\sim 10^{-2}$ Jy. 
\par

\section{Discussion}\label{DIS}
An electron population with an energy distribution steeper than $\gamma_{\rm e}^2$ is required to generate plasma maser radiations. 
\cite{1984A&A...136..227S} developed an evolutionary model for relativistic electron energy spectra based on Fermi acceleration at shock fronts, second-order Fermi acceleration due to moving magnetized fluid elements, and energy losses due to synchrotron and inverse Compton radiation. Regardless of the injection spectrum of the particles, the energy distribution of the accelerated particles exhibits a power law form with a peak at $\gamma_{\rm e,s}$, where $\gamma_{\rm e,s}$ corresponds to the point at which the Fermi acceleration timescale equals the radiation loss timescale. The spectral index of the power law below $\gamma_{\rm e,s}$ is determined by the ratio of the acceleration timescale to the particle escape lifetime. In the case of a very long escape lifetime, particles pile up sharply around $\gamma_{\rm e,s}$, and the spectrum exhibits an exponential cutoff beyond $\gamma_{\rm e,s}$, resulting in an almost monoenergetic particle distribution. In addition, collisionless shocks can result in a nonthermal component with $ n_{\rm e} \propto \gamma_{\rm e}^{-p}$ in the high-energy regime, and a thermal component with $\gamma_{\rm e}^2$ in the low-energy regime \citep{1986rpa..book.....R,2008ApJ...682L...5S}. These two components are joined at $\gamma_{\rm e,s}$. As nonthermal electrons cool radiatively, they gradually lose energy and accumulate at lower energies near $\gamma_{\rm e,s}$, forming a very narrow distribution at $\gamma_{\rm e,s}$ \citep{2002ApJ...574..861S}. Therefore, in our calculations of plasma maser emission, we simply approximated the electron distribution as a monoenergetic population.

\begin{figure}
\includegraphics[width=1\columnwidth]{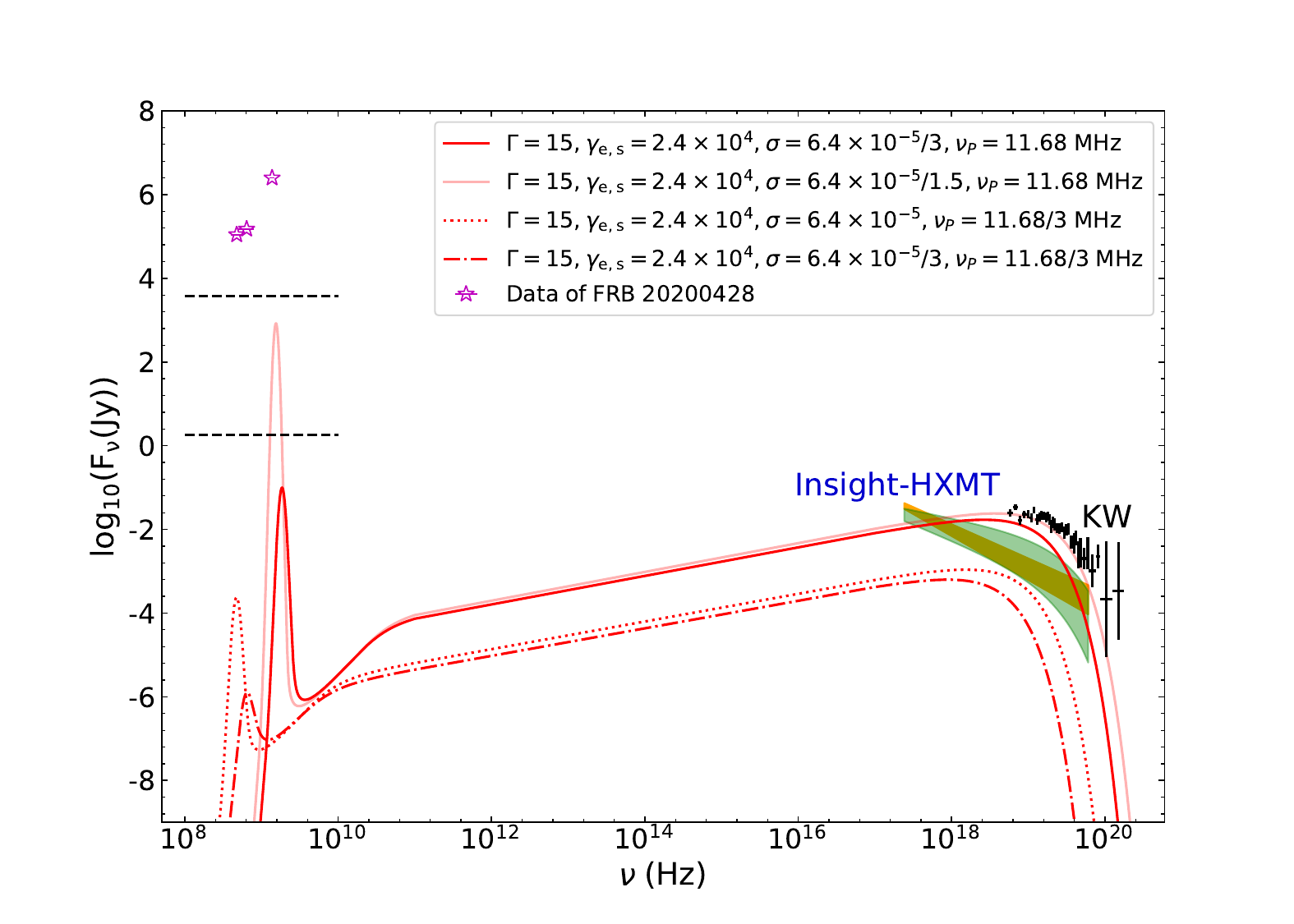}
 \centering
\caption{Same as Fig. \ref{MyFig1} but the model parameters are varied, as marked in the panel, to illustrate a possible explanation for the observed sub-energetic radio bursts or giant radio pulses from SGR 1935+2154. The dashed black lines indicate the range of several bright radio bursts detected in follow-up observations of SGR 1935+2154.}\label{MyFig3}
\end{figure}

The peak flux density of plasma maser emission is sensitive to the $\sigma$ and $\nu_P$ of a plasma blob. Figure \ref{MyFig3} shows the spectra of the plasma maser emission and synchrotron emission when adopting different sets of $\sigma$ and $\nu_P$ values. One can observe that the peak flux density of the maser emission varies by more than 10 orders of magnitude, while the flux density of the synchrotron emission only varies by $1-2$ orders of magnitude. This presents a potential explanation for the observed sub-energetic radio bursts or giant radio pulses from SGR 1935+2154. We note that follow-up observations of SGR 1935+2154 detected several bright radio bursts with peak flux densities ranging from $1.8$ Jy to $3.8$ kJy \citep{2021NatAs...5..414K,2023arXiv231016932G}. One of them was detected simultaneously with a short X-ray burst observed by the Gravitational-wave high-energy electromagnetic counterpart all-sky monitor (GECAM), Insight-HXMT high-energy burst searcher (HEBS), and KW. This is consistent with our model prediction.

\par

The size and location of the radiating blobs are of interest. Based on the derived parameter sets from our Monte Carlo simulation analysis, we inferred the size of the blobs to be $\sim 10^{9-10}$ cm and the magnetic field of the plasma blobs to be $B=10^2\sim 0.7\times 10^4$ G. Assuming that FRBs are powered by the magnetic energy of a magnetar \citep{2014MNRAS.442L...9L}, we estimated the location of the FRB radiating blob to be $r_{\rm FRB}\sim b{B_{\rm p}}{R_{\rm p}}/B$, where $R_{\rm p}$ is the radius of the magnetar and $b\;(< 1)$ is a dimensionless constant. SGR 1935+2154 has a spin period of $P \simeq 3.24$ s, a spin-down rate of $\dot P \simeq 1.43 \times {10^{ - 11}}\;\rm s\;{s^{ - 1}}$, and a surface dipole magnetic field strength of ${B_{\rm p}} \simeq 2.2 \times {10^{14}}$ G
\citep{2014GCN.16533....1G,2016MNRAS.457.3448I}. We estimated the value of $b$ as $b\sim (E_{\rm XRB}/E_{\rm B})^{1/2}$ \citep{2020ApJ...899L..27M}, where $E_{\rm B}$ is the total magnetic energy of SGR 1935+2154 and $E_{\rm XRB}$ is the released energy of the two narrow X-ray peak bursts (which lasted 4 ms) associated with FRB 20200428. We have $E_{\rm B}={B_{\rm p}^{2} R_{\rm p,6}^{3}}/{6} \sim 10^{46}$ erg. The value of $E_{\rm XRB}$ for the hard X-ray peak bursts from SGR 1935+2154 is $\sim 10^{38}$ erg \citep{2021NatAs...5..372R}. Thus, we have $b\sim 10^{-4}$. 
We obtain ${r_{\rm FRB}}= 3.1\times 10^{12}\sim2.2\times10^{14}$ cm, which suggests that the sites of the FRB bursts are far from the central engine. This is evidence in support of the ``far-away'' scenarios \citep{2020MNRAS.498.1397L,2020ApJ...899L..27M,2021MNRAS.500.2704Y}. 
\par

It should also be noted that the fitting results shown in Fig. \ref{MyFig1} correspond to the time-integrated spectrum of the $\sim$ 1 s X-ray data observed by Insight-HXMT during the FRB 20200428 burst \citep{2021NatAs...5..378L}, as well as the X-ray data with a duration of $\sim 0.46$ s observed by KW \citep{2021NatAs...5..372R}. However, this comparison is reasonable because the hardness of the light curve of the X-ray burst reaches its maximum at the 
location of the two narrow peaks and lasts $\sim1$ ms. This means that the two narrow peaks must be dominated by a nonthermal spectrum \citep{2021NatAs...5..378L,2021NatAs...5..372R}. In addition, the non-detection of radio emission from about one hundred typical soft bursts of SGR 1935+2154 further supports the idea that magnetar FRB signals are associated with rare X-ray bursts with hard spectra \citep{2020Natur.587...63L}.
\par

\section{Conclusions}\label{CON}
We have demonstrated that the observed flux densities and
spectra of both FRB 20200428 and two potentially associated narrow X-ray peak bursts from SGR 1935+2154 can be explained using our model, which attributes the FRB and the X-ray bursts to plasma synchrotron maser emission and synchrotron radiation from weakly magnetized relativistic plasma blobs, respectively. We performed a Monte Carlo simulation analysis to constrain the model parameters that can generate observable FRB outbursts and potential X-ray bursts. It is found that $\Gamma$ is in the range $5-30$. The $\gamma_{\rm e,s}$ is uniformly distributed in the range $1.8\times10^4-3.3\times10^4$. The distribution of $\sigma$ exhibits three separated peaks, at $\rm log(\sigma)=\left\{-4.21,-3.95,-3.88\right\}$, corresponding  to the three detected bursts of FRB 20200428. The distribution of $\nu_P$ ranges from $\log \nu_{P}=6.40$ ($\nu_{P}=2.48 \,{\rm MHz}$) to $\log \nu_{ P}=7.6$ ($\nu_{ P}=42.61\, {\rm MHz}$). The inferred size of the blobs is $\sim 10^{9-10}$ cm, and the blob is located $\sim 10^{12-14}$ cm from the central engine. Adopting fine-tuned parameter sets as $\{{{\Gamma,\gamma _{{\rm{e,s}} }},\sigma ,{\nu_{ P}/{\rm MHz}} }\}=\{15,2.4\times10^4,6.4\times10^{-5},11.68\}, \{15,2.4\times10^4,13.3\times10^{-5},6.48 \}$, and $\{15,2.4\times10^4,17.9\times10^{-5},5.23\}$, the observed spectra of both FRB 20200428 and the X-ray bursts from SGR 1935+2154 can indeed be reproduced by our model. Furthermore, the peak flux density of plasma maser emission is sensitive to $\sigma$ and $\nu_P$. It can vary by over 10 orders of magnitude, whereas the synchrotron emission flux density only varies by $1-2$ orders of magnitude, consistent with the observed sub-energetic radio bursts or giant radio pulses from SGR 1935+2154.
\par

As illustrated in Fig. \ref{MyFig1}, the synchrotron emission spectrum of plasma blobs is in the range $10^{10}$ to $\sim 10^{18}$ Hz. This suggests that short-duration optical and X-ray flashes can be accompanied by FRB bursts. Recently, an extragalactic FRB source, FRB 20250316A, was reported to be possibly associated with the X-ray source EP J120944.2+585060 detected by the Einstein Probe \citep{2025ATel17100....1S}. However, follow-up observations with the \textit{Chandra} X-ray telescope did not support this association \citep{2025ATel17119....1S}. FRB 20250316A is a bright FRB from the nearby galaxy NGC 4141 and is located at a distance of $\sim$ 40 Mpc \citep{2025ATel17081....1N}. The nearest known extragalactic FRB is FRB 20200120E, which is located in a globular cluster of the M81 galaxy at a distance of $\sim$ 3.6 Mpc \citep{2021ApJ...910L..18B,2022Natur.602..585K,2024NatCo..15.7454Z}. In contrast, typical FRBs are found at distances of $\sim$ 1 Gpc. If FRB 20200428 had occurred at a distance of 10 Mpc, its radio flux density would be $\sim$ 1 Jy (${F_\nu } \propto \rm D_L^{ - 2}$), while the X-ray and multiwavelength counterpart would be too faint to be detectable. Searching for associated X-ray/optical counterparts of cosmic FRBs is challenging with current time-domain astronomy. 

\begin{acknowledgements}
We thank the anonymous referee for helpful comments. 
We thank the helpful discussions with Shu-Qing Zhong, Hao-Hao Chen and Ying Gu. This work is supported by the National Key R\&D Program (2024YFA1611700) and the National Natural Science Foundation of China (grant Nos. 12133003). E. W. L. is supported by the Guangxi Talent Program (``Highland of Innovation Talents''). 
\end{acknowledgements}

\bibliographystyle{aa} 
\bibliography{example} 

\end{document}